\documentclass{aastex61}
\pdfoutput=1 
\usepackage{amsmath,amstext}
\usepackage[T1]{fontenc}
\usepackage{apjfonts} 
\usepackage[figure,figure*]{hypcap}
\usepackage{todonotes}
\allowdisplaybreaks

\shorttitle{A Fast-Converging Digital Adaptive Filter}
\shortauthors{Finger R. et al.}

\begin{document}

\title{A FPGA-Based Fast Converging Digital Adaptive Filter for Real-Time RFI mitigation on Ground Based Radio Telescopes}

\author{R. Finger}
\affiliation{Department of Astronomy, Universidad de Chile}
\affiliation{Department of Electrical Engineering, Universidad de Chile}
\author{F. Curotto}
\affiliation{Department of Astronomy, Universidad de Chile}
\affiliation{Department of Electrical Engineering, Universidad de Chile}
\author{R. Fuentes}
\affiliation{Department of Astronomy, Universidad de Chile}
\affiliation{Department of Electrical Engineering, Universidad de Chile}
\author{R. Duan}
\affiliation{National Astronomical Observatories, Chinese Academy of Sciences}
\author{L. Bronfman}
\affiliation{Department of Astronomy, Universidad de Chile}
\author{D. Li}
\affiliation{National Astronomical Observatories, Chinese Academy of Sciences}
\affiliation{Key Laboratory of Radio Astronomy, Chinese Academy of Sciences} 

\begin{abstract}
Radio Frequency Interference (RFI) is a growing concern in the radio astronomy community. Single dish telescopes are particularly susceptible to RFI. Several methods have been developed to cope with RF-polluted environments, based on flagging, excision, and real-time blanking, among others. All these methods produce some degree of data loss or require assumptions to be made on the astronomical signal.
We report the development of a real-time, digital adaptive filter implemented on a Field Programmable Gate Array (FPGA) capable of processing 4096 spectral channels in a 1 GHz of instantaneous bandwidth. The filter is able to cancel a broad range of interference signals and quickly adapt to changes on the RFI source, minimizing the data loss without any assumption on the astronomical or interfering signal properties. The speed of convergence (for a decrease to a 1\%) was measured to be 208.1 $\mu s$ for a broadband noise-like RFI signal and 125.5 $\mu s$ for a multiple-carrier RFI signal recorded at the FAST radio telescope.

\end{abstract}


\section{Introduction}
Radio Frequency Interference (RFI) is a growing concern in the radio astronomy community. Wireless communication devices have never been more ubiquitous, and sensing devices such as automotive radars for automation (cruise control) and safety (pre-collision airbag deployment) are using high frequency bands not used for commercial applications before. Nowadays we can find consumer applications using frequencies from a few KHz to 81 GHz and beyond, most of them on mobile devices \citep{hasch2015driving}. The radio astronomy community is well aware of the problem and has developed a number of different methods to cope with various types of RFI (refer to \citep{series_techniques_2013} for a comprehensive list). Interferometers have additional capabilities to mitigate RFI sources by synthesizing a beam null over the source, A technique called null-steering \citep{fridman_radio_2005}, but it requires knowledge of the source position in time.

Single dish telescopes are particularly susceptible to RFI coupling through antenna sidelobes. For single dish telescopes several methods have been described in literature based on flagging and excision, real-time blanking, and frequency flagging \citep{nita_radio_2007,winkel_rfi_2007,niamsuwan_examination_2005,offringa_post-correlation_2010}. All these methods produce some degree of data loss or require assumptions to be made on the astronomical signal. Currently, ad-hoc methods are used for each type of RFI, and no general solution has been adopted to cope with the problem.

A more ambitious approach is to identify, characterize, and subtract the RFI from the astronomical signal in real-time by using adaptive filters \citep{barnbaum_new_1998,briggs_removing_2000,kesteven2004tc}. The main pro of this approach is that it could deal with RFI with burst and persistent characteristics, broad or narrow band, and tackle fast moving sources or interference with unknown direction of arrival, provided that the adaptive filter converges in time scales shorter than the changes of the RFI. Nevertheless no reports on the speed of convergence of such adaptive filters are found in literature, and only modest bandwidths have been reported.

Here we report the development of a real-time, frequency domain digital adaptive filter implemented on a state of the art Field Programmable Gate Array (FPGA) capable of processing 4096 spectral channels in a 1 GHz of instantaneous bandwidth. The filter is able to cancel a broad range of RFI types and quickly adapt to changes on the RFI source characteristics, minimizing the data loss without any assumption on the astronomical or interfering signal properties.

\section{Filter Design}
The filter architecture is based on the work of \citep{kesteven_rfi_2010}. The design assumes that the RFI signal can be independently sampled without overlap of the astronomical signal. Specifically, it is assumed that the astronomical signal and the sampled RFI are uncorrelated. Since RFI is normally many orders of magnitude stronger than the astronomical signals, and the reference antenna can be made to have high gain over the horizon with almost no gain on the sky, the above assumption is easy to meet for most astronomical sources and RFI environments. Nevertheless, moving, high elevation RFI sources might require a tracking directional reference antenna.

\begin{figure}[ht!]
\centering
\includegraphics[width=0.7\textwidth]{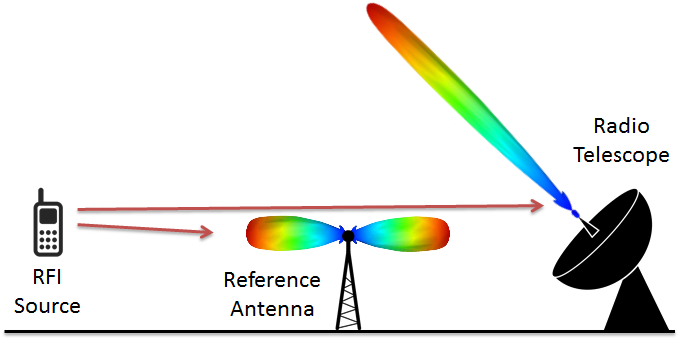}
\caption{Radio telescope (primary) and reference antenna radiation patterns examples. The RFI couples to the primary and reference antenna, but the astronomical signal is only detected by the radio telescope.}
\label{fig:example}
\end{figure}

To sample and subtract the RFI, both the primary and reference signals have to be digitized by two Analog to Digital Converters (ADCs) sharing the same clock source, i.e. with a couple of synchronized ADCs. After digitalization, two pipelined Fast Fourier Transforms (FFTs) of 4096 spectral bins are calculated. Synchronized ADCs and FFTs have been implemented for digital signal processing in radio astronomy, particularly applied to digital sideband separation \citep{finger2013calibrated,finger2015ultra}. In our design the ADCs run at 2 GSPS, to sample a DC to 1 GHz baseband.

\begin{figure}[ht!]
\centering
\includegraphics[width=\textwidth]{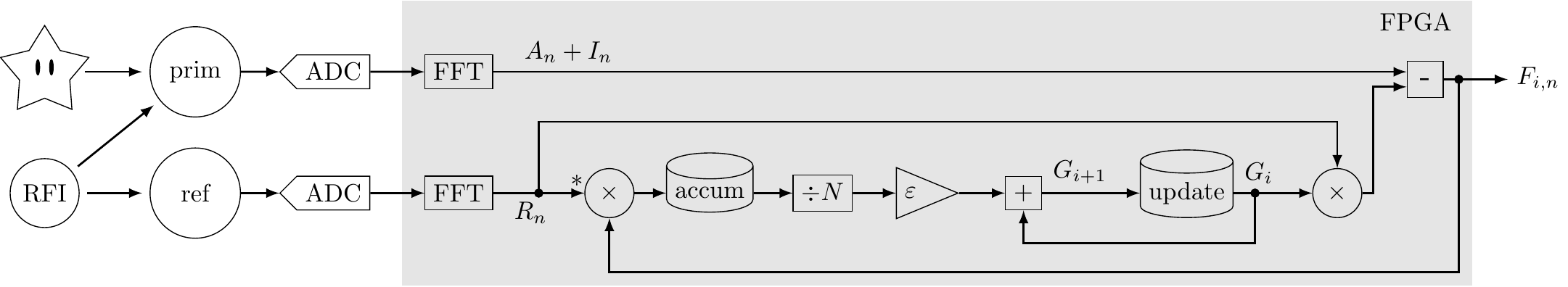}
\caption{Digital adaptive filter block diagram.}
\label{fig:algorithm}
\end{figure}

We use the following convention to name the different signals within the system:

\begin{description}
\item[$A_n$] Astronomical signal frequency sample $n$.
\item[$I_n$] Interference signal frequency sample $n$.
\item[$R_n$] Interference signal measured in the reference antenna.
\item[$G_i$] Update output at cycle $i$ (after $N$ samples accumulations).
\item[$F_{i,n}$] Filter output sample $n$ after $i$ update cycles
\end{description}

Figure~\ref{fig:algorithm} shows the block diagram of the filter.

After digitalization two pipelined FFTs of 4096 spectral bins are calculated. The reference channel is then conjugated and multiplied by the output signal. The filter itself has two sequential cycles: accumulation and update. During the accumulation cycle, $N$ samples are added while the update register keeps the value calculated in the last accumulation cycle. The accumulation length $N$ is a user selectable parameter. After each accumulation cycle, $G_i$ is updated to be the sum of its last value plus the accumulator output, scaled by the constant $\varepsilon/N$. The parameter $\varepsilon$ is also user selectable. The update of $G_i$, in turn, updates the filter's output. Finally the filter output is integrated for 536 $ms$ to mimic the operation of an astronomical back-end.

For the mathematical analysis, it is assumed that the primary and the reference signals are purely sinusoidal, of constant amplitude and phase, which in the frequency domain can be represented by complex values.

With the above definitions we can write the following differences equations:

\begin{align}
G_0&=0 \label{eq:init1}\\
G_{i+1}&=\varepsilon\frac{1}{N}\sum_{j=Ni}^{N(i+1)-1}F_{i,j}R_j^*+G_i \label{eq:init2}\\
F_{i,n}&=A_n+I_n-G_iR_n \label{eq:init3}
\end{align}

Where~\eqref{eq:init1} is the initial condition of the filter, \eqref{eq:init2} and \eqref{eq:init3} are the update and the filter output equations. To solve this system of equations we first replace \eqref{eq:init3} in \eqref{eq:init2}.

\begin{align}
G_{i+1}&=\varepsilon\frac{1}{N}\sum_{j=Ni}^{N(i+1)-1}\left(A_j+I_j-G_iR_j\right)R_j^*+G_i\\
&=\varepsilon\frac{1}{N}\sum_{j=Ni}^{N(i+1)-1}A_jR_j^*+I_jR_j^*-G_iR_jR_j^*+G_i\\
&=\varepsilon\frac{1}{N}\underbrace{\sum_{j=Ni}^{N(i+1)-1}A_jR_j^*}_{=C_i}+\varepsilon\frac{1}{N}\sum_{j=Ni}^{N(i+1)-1}\underbrace{I_jR_j^*}_{K}-\varepsilon\frac{1}{N}G_i\sum_{j=Ni}^{N(i+1)-1}\underbrace{R_jR_j^*}_{r}+G_i \label{eq:expand}\\
&=\varepsilon\frac{1}{N}C_i+\varepsilon\frac{1}{N}\sum_{j=Ni}^{N(i+1)-1}K-\varepsilon\frac{1}{N}G_i\sum_{j=Ni}^{N(i+1)-1}r+G_i\\
&= G_i - \varepsilon r G_i + \frac{\varepsilon}{N}C_i + \varepsilon K\\ 
&=\underbrace{(1-\varepsilon r)}_{:=b}G_i + \frac{\varepsilon}{N}C_i + \varepsilon K\label{eq:def_b}\\
G_{i+1}&=bG_i + \frac{\varepsilon}{N}C_i + \varepsilon K \label{eq:diff}
\end{align}

Where in \eqref{eq:expand} we use the fact that $I_n$ and $R_n$ are correlated by $K$. In~\eqref{eq:def_b} we assume that $r$ changes slowly compared to the filter update cycle $i$, so we can define
 $b:=1-\varepsilon r$. This yields a first order difference equation in~\eqref{eq:diff}; its solution is given by:

\begin{align}
G_i&=\frac{\varepsilon}{N}\sum_{k=0}^{i-1}b^kC_{i-k-1}+\varepsilon K\frac{1-b^i}{1-b}\\
&=\frac{\varepsilon}{N}\sum_{k=0}^{i-1}b^kC_{i-k-1}+\frac{K}{r}(1-b^i)
\end{align}

Choosing $\varepsilon$ such that $|b|<1$ ensures the filter converges. Replacing the obtained expression for $G_i$ in~\eqref{eq:init3} we get:
\begin{align}
F_{i,n}&=A_n+I_n-R_n\left(\frac{\varepsilon}{N}\sum_{k=0}^{i-1}b^kC_{i-k-1}+\frac{K}{r}(1-b^i)\right)\\
&=A_n+I_n-R_n\frac{K}{r}(1-b^i)-\frac{\varepsilon}{N}R_n\sum_{k=0}^{i-1}b^kC_{i-k-1}\\
&=A_n+I_n-\frac{R_nR_n^*}{r}(1-b^i)I_n-\frac{\varepsilon}{N}R_n\sum_{k=0}^{i-1}b^kC_{i-k-1}\\
&=A_n+b^iI_n-\frac{\varepsilon}{N}R_n\sum_{k=0}^{i-1}b^kC_{i-k-1} \label{eq:output}
\end{align}
From~\eqref{eq:output} it can be concluded that for $i\rightarrow\infty$ the interference $I_n$ gets removed, as long as the filter converges faster than the change of the RFI. The speed of the filter is controlled by $\varepsilon$, where $b\approx 0$ (i.e. $\varepsilon\approx 1/r$), yields the fastest convergence. The third term represent the filter added noise. The noise term scales with $\varepsilon/N$, so added noise can be reduced at expenses of convergence speed by increasing $N$.

Notice that no assumption was made about the signals frequency so this monochromatic analysis can be extended to any number of spectral bins. Implementing one filter per each spectral channel allows processing any type of broad or narrow band signal.


\section{Filter implementation and test setup}

The described adaptive filter was implemented in a ROACH2\footnote{\url{https://casper.berkeley.edu/wiki/ROACH2}} (Reconfigurable Open Architecture Computing Hardware) development board designed by the CASPER group (Collaboration for Astronomy Signal Processing and Electronics Research) led by Berkeley University. The most important components of the board are the Xilinx Virtex-6\footnote{\url{https://www.xilinx.com/support/documentation/data_sheets/ds150.pdf}} XC6VSX475T FPGA, and a couple of 5GSPS ADC boards \citep{jiang20145} manufactured by ASIAA (Academia Sinica Institute of Astronomy and Astrophysics), which use the e2v EV8AQ160\footnote{\url{https://www.e2v.com/resources/account/download-datasheet/2291}} ADC chips. The FPGA is programmed via the MATLAB Simulink software, using specialized libraries provided by Xilinx and CASPER group \citep{parsons_new_2006}.

\begin{figure}[ht!]
\centering
\includegraphics[width=0.66\textwidth]{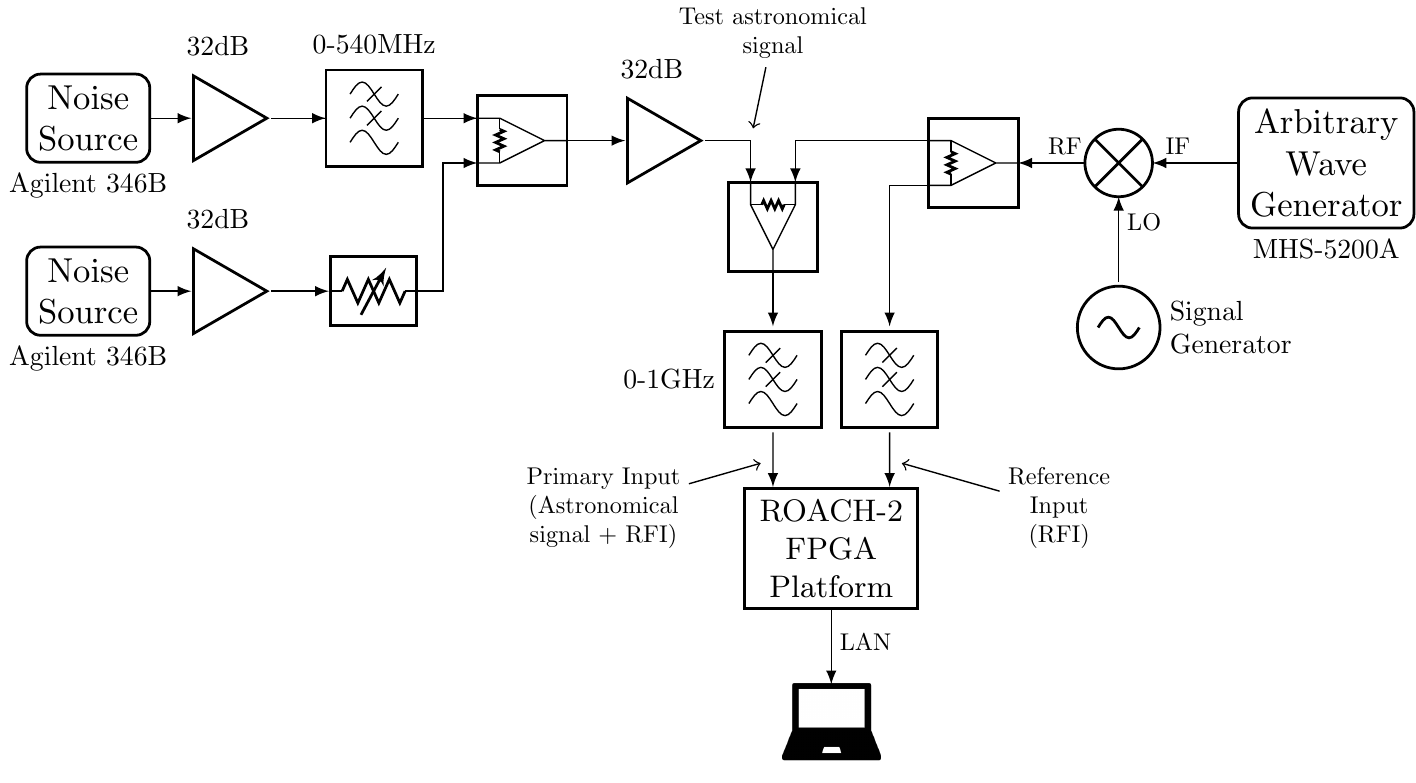}
\caption{Test setup Block Diagram. At the left a noise diode is used as test astronomical signal. At the right is depicted the Arbitrary Waveform Generator and up-converter stage used as RFI source}
\label{fig:setup}
\end{figure}

A test setup was assembled to characterize the adaptive filter performance with different types of RFI. A band limited noise signal was used as astronomical source, and an Arbitrary Waveform Generator (AWG) with an up-conversion stage was used as RFI source. Figure~\ref{fig:setup} shows the test setup block diagram.

\begin{figure}[ht!]
\plottwo{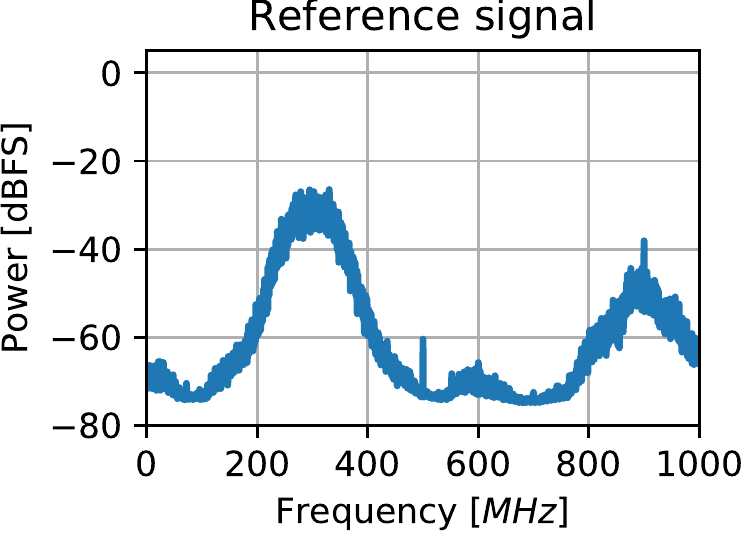}{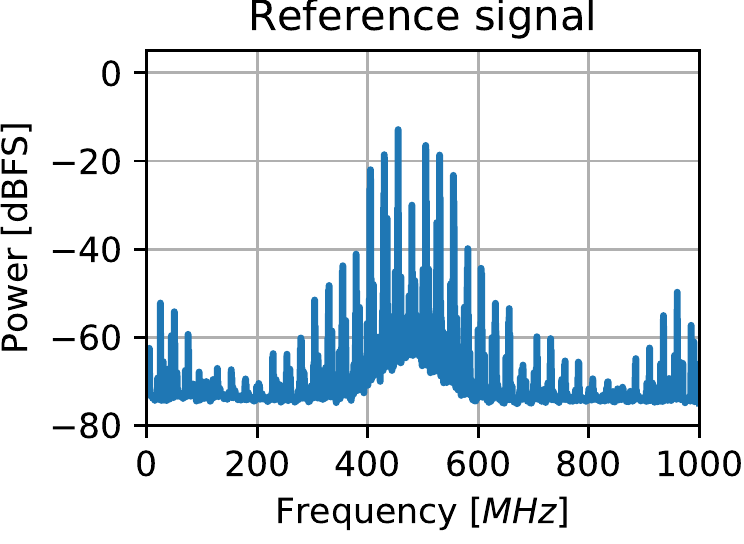}
\caption{Noise-like RFI (left) and multiple-carrier RFI (right) test signals. The line at 500 MHz in the left panel is a spurious signal coming from the interleaving of the ADC cores. The line at 900 MHz is the third harmonic of the local oscillator used to up-convert the AWG output}
\label{fig:interference}
\end{figure}

\begin{figure}[ht!]
\centering
\includegraphics[width=0.5\textwidth]{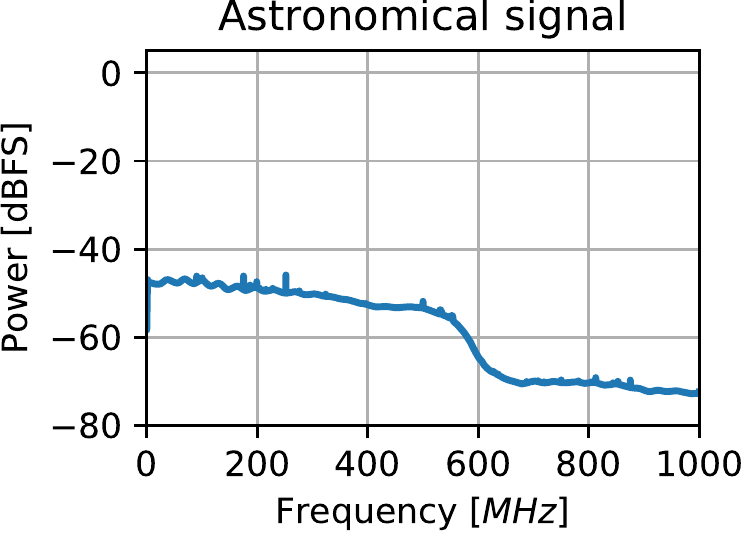}
\caption{Band limited noise used as the astronomical signal.}
\label{fig:astronomical}
\end{figure}

Two types of RFI were used to test the filter: A broadband noise-like signal and a stronger and complex multiple-carrier signal. The broadband signal was generated via software while the multiple-carrier signal was produced from RFI measured at the Five Hundred Meter Aperture Spherical Telescope (FAST, \citep{nan2011five}) site. The RFI at FAST was recorded with a reference antenna mounted on top of one of the towers that holds the feed horn cabin.
 Figure~\ref{fig:interference} shows the RFI test signals while Figure~\ref{fig:astronomical} depicts the astronomical test signal.

\pagebreak

\section{Results}

\begin{figure}[ht!]
\centering
\includegraphics[height=5.5cm]{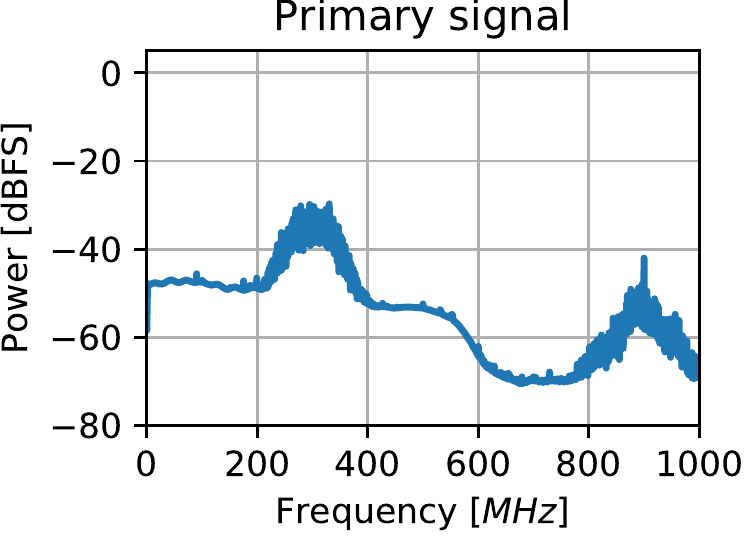}
\includegraphics[height=5.5cm]{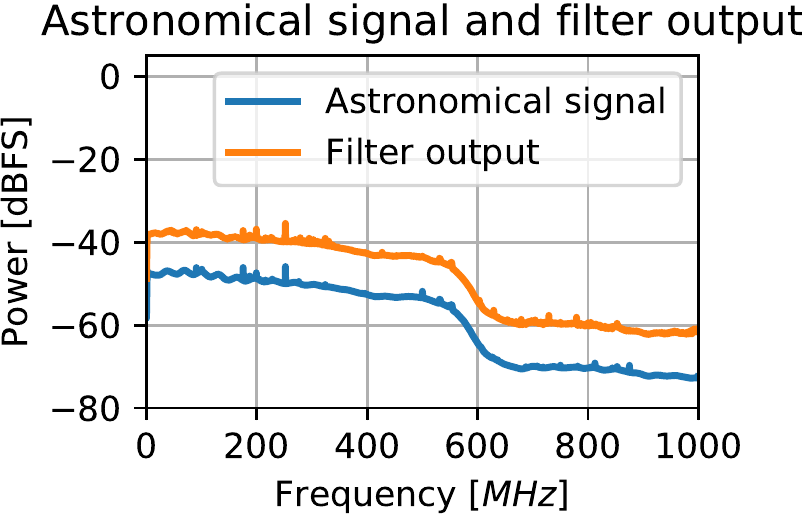}
\caption{Astronomical signal contaminated with broadband RFI (left) and the filtered output (right). The original astronomical signal is also plotted (right) for comparison. The output was offset by 10dB for better visualization. The parameters of the filter are $\varepsilon=2^{15}$, $N=1$}
\label{fig:res_noise}
\end{figure}

\begin{figure}[ht!]
\centering
\includegraphics[height=5.5cm]{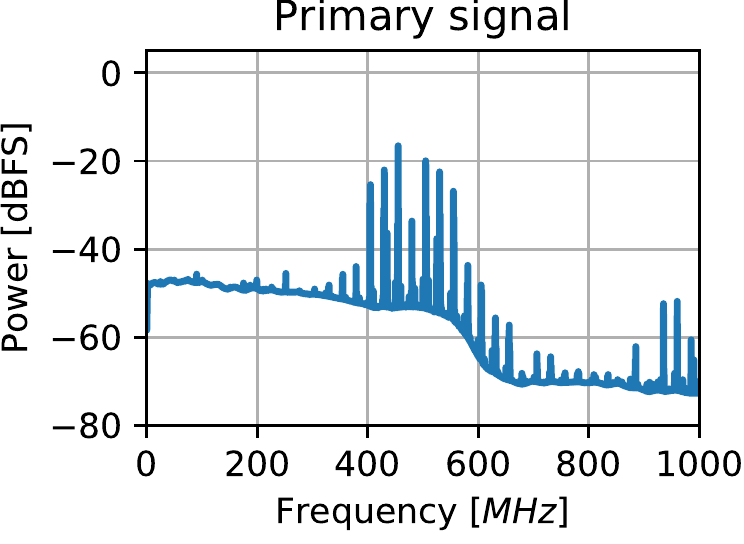}
\includegraphics[height=5.5cm]{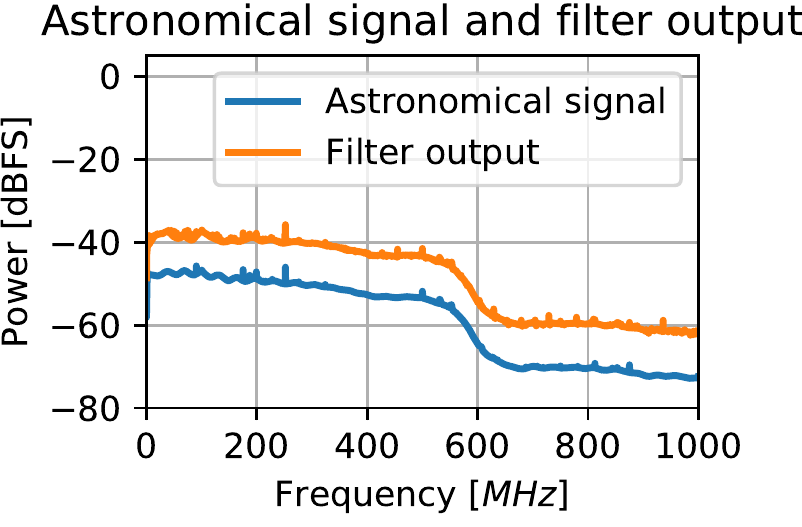}
\caption{Astronomical signal contaminated with multi-carrier RFI measured at FAST site (left) and the filtered output (right). The output was offset by 10dB for better visualization. The parameters of the filter are $\varepsilon=2^{11}$, $N=1$}
\label{fig:res_fast}
\end{figure}

Figures~\ref{fig:res_noise} and \ref{fig:res_fast} show the astronomical signal contaminated with RFI and the filtered output. No sign of the RFI can be seen in the filtered output for both broadband and multi-carrier RFI. For the filter parameters it was chosen $N=1$ (no accumulation in the filter) in order to achieve maximum convergence speed. For $\varepsilon$ it was given a value that roughly satisfy $\varepsilon r_{max}\approx 1$, where $r_{max}$ is the maximum power among all channels of the reference signal, which is different between each type of interference: $2^{15}$ for the Multiple-carrier RFI and $2^{11}$ the broadband RFI.

\begin{figure}[ht!]
\plottwo{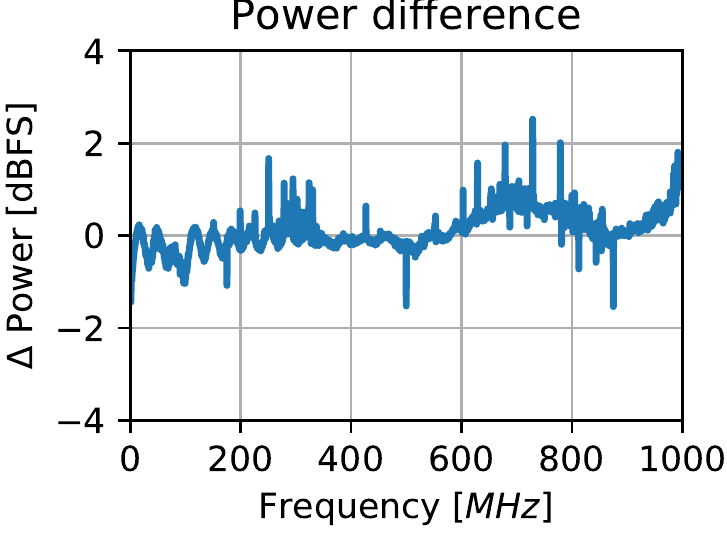}{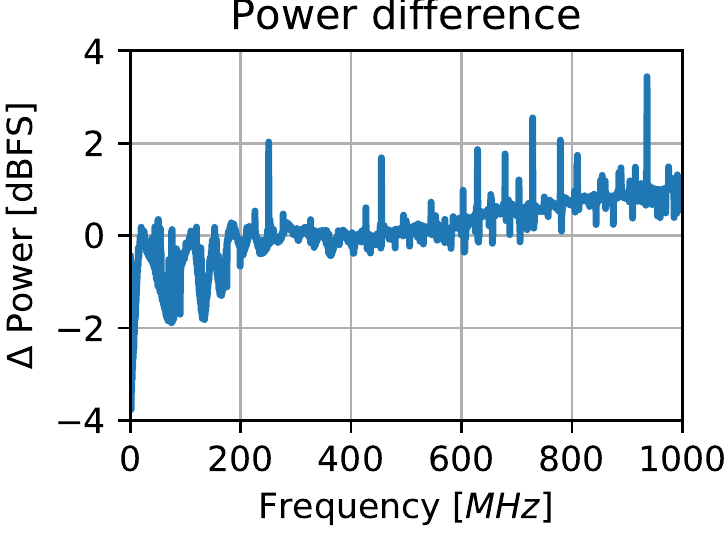}
\caption{Difference between astronomical signal and filter output, for broadband RFI (left), and RFI recorded from FAST (right).}
\label{fig:diff}
\end{figure}

Figure~\ref{fig:diff} shows the difference between the astronomical signal and the filter output. The difference is in general within +/- 1 dB, except for specific channels where differences above 2dB can be seen. These channels are correlated with ADC spurious signals.
The filter output total power is 4\% less than the astronomical signal for the broadband RFI and 9\% less for the FAST RFI.  


\begin{figure}[ht!]
\plottwo{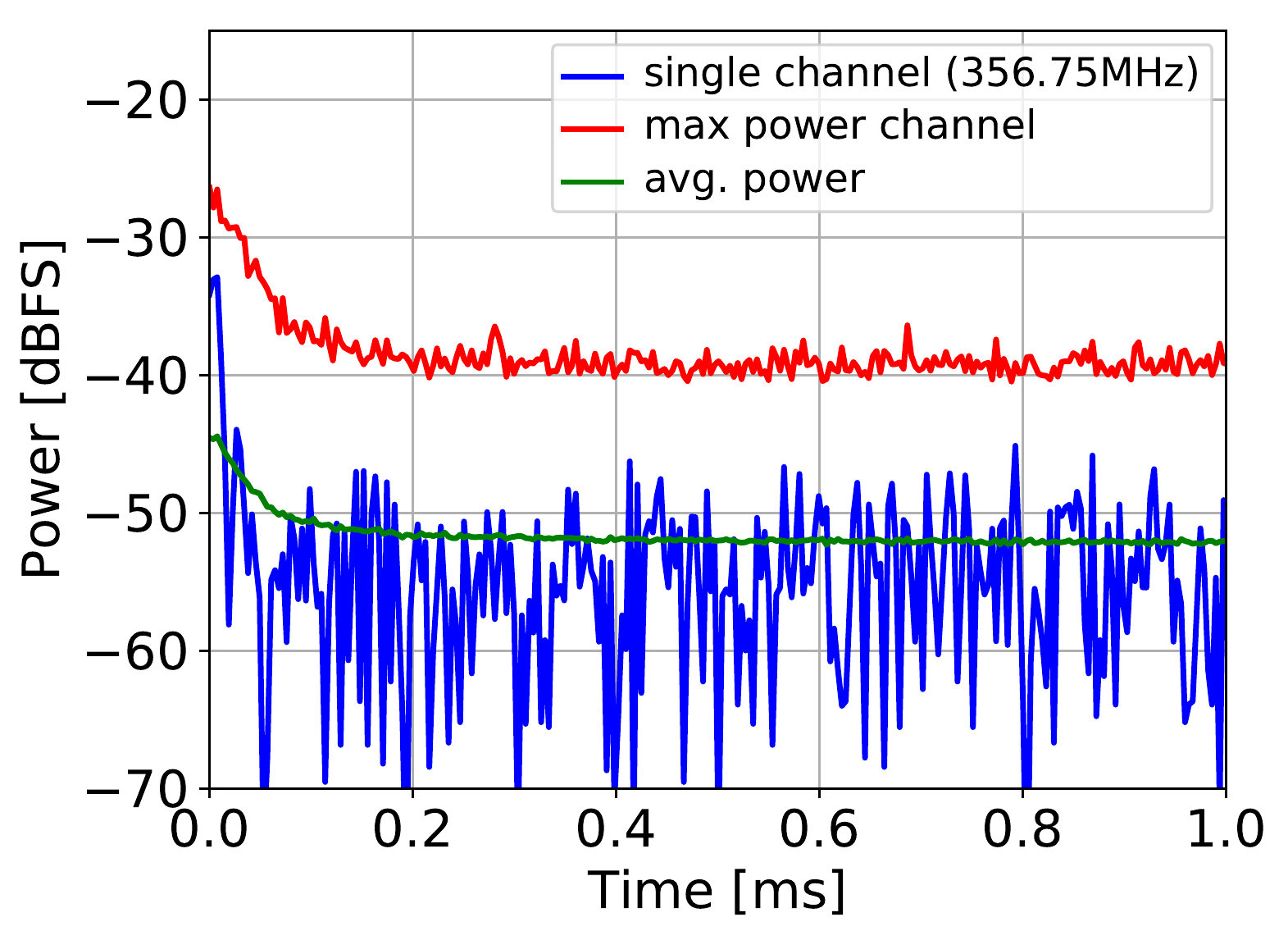}{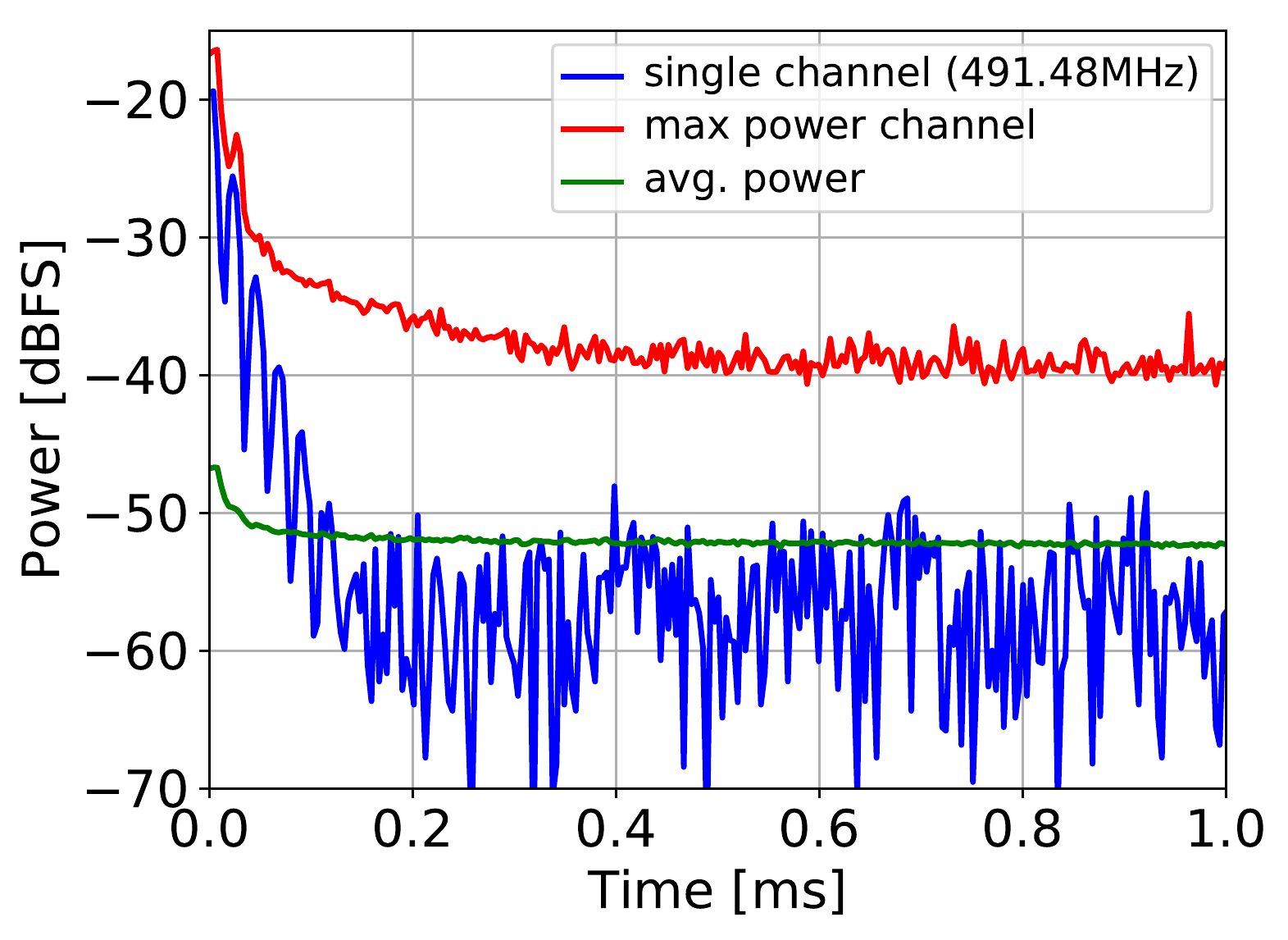}
\caption{Output power v/s time for broadband RFI (left), and multiple-carrier RFI recorded at FAST (right). Three ways of measuring convergence are depicted: The value of the maximum spectral channel at $T=0$ (blue line), the maximum spectral channel at each time (red line), and the average power of the whole spectrum (green line).}
\label{fig:convergence}
\end{figure}

Figure~\ref{fig:convergence} shows the speed of convergence of the filter. $T=0$ is defined as the moment the update loop is closed with a $G_0=0$. Convergence, measured for the average power (green) line, as the time needed to fall to a 1/e (36.8 \%) of the $T=0$ level, is reached on 45.2 $\mu s$ for the broadband RFI and 27.2 $\mu s$ for the multiple-carrier RFI. Convergence to a 1\% level is reached in 208.1 $\mu s$ for the broadband RFI and 125.5 $\mu s$ for the multiple-carrier RFI.



\pagebreak
~
\pagebreak

\section{Conclusions}
\begin{itemize}
\item A fast-converging adaptive filter was implemented on a FPGA achieving the real time processing of 4096 spectral channels on a 1 GHz bandwidth.
\item Using no assumptions on the astronomical signal properties, it was analytically demonstrated that the filter converges to an output clean of RFI.
\item Experiments were done in laboratory using broadband and narrowband RFI test signals. In both cases the output equals the clean astronomical signal with negligible residual contamination.
\item The time of convergence (for a decrease to a 1\%) was 208.1 $\mu s$ for a broadband noise-like RFI signal and 125.5 $\mu s$ for a multiple-carrier RFI signal recorded at the FAST radio telescope, demonstrating that the filter can adapt to very fast-changing RFI, like the one produced by burst transmissions from fast moving sources.
\item Embedded in an astronomical back end, the reported filter should be able to cope with broad and narrow band RFI with negligible data loss.
\item The adaptive filter may also find applications in communications on highly RF-polluted environments, and in defense, to counteract GPS or radio communication jammers.
\end{itemize}

\acknowledgments
This work was supported by the Chilean National Commission for Scientific and Technological Research (CONICYT) through its grants CATA-Basal PFB06, ALMA 31150012 and FONDECYT 11140428, and by the Chinese Academy of Science South America Center for Astronomy (CASSACA). DL and DR also acknowledge the support from the International Partnership Program of Chinese Academy of Sciences, Grant No.114A11KYSB20160008. 
We thank Xilinx Inc. for the donation of FPGA chips and software licenses.


\bibliographystyle{yahapj}
\bibliography{references}


\end{document}